\documentclass[aps,prb,twocolumn,superscriptaddress,nofootinbib,reprint]{revtex4-2}
\usepackage{makeidx}
\usepackage{graphicx}
\usepackage{amsmath}
\usepackage{amsfonts}
\usepackage{amssymb}
\usepackage{bm}
\usepackage{xcolor}
\usepackage{float}
\usepackage{physics}
\usepackage{upgreek}
\usepackage[version=3]{mhchem}
\usepackage{caption}

\begin{document}
\title {Magnon-photon coupling in the YIG-based disk and ring microcavities}

\author{S.S. Demirchyan}
\affiliation{Russian Quantum Center, 30 Bolshoy Boulevard, building 1, Skolkovo Innovation Center territory, Moscow, 121205, Russia}

\author{D.M. Krichevsky}
\email{dm.krichevsky@physics.msu.ru}
\affiliation{Russian Quantum Center, 30 Bolshoy Boulevard, building 1, Skolkovo Innovation Center territory, Moscow, 121205, Russia}
\affiliation{Faculty of Physics, Lomonosov Moscow State University, Leninskie Gory, Moscow 119991, Russia}

\author{V.I. Belotelov}
\affiliation{Russian Quantum Center, 30 Bolshoy Boulevard, building 1, Skolkovo Innovation Center territory, Moscow, 121205, Russia}
\affiliation{Faculty of Physics, Lomonosov Moscow State University, Leninskie Gory, Moscow 119991, Russia}

\begin{abstract}
Optomagnonic dielectric resonators offer a promising platform for the bidirectional conversion of microwave and optical photons at the single quantum level. Current implementation of such a conversion lacks from low magneto-optical interaction strength, limiting its practical utilization in quantum technologies. The main bottleneck is the small spatial overlap between optical and magnon modes. Here, we show that utilization of a disk and ring geometries notably increases the mode overlap. We analyze the interaction volume of optical whispering gallery and magnon Kittel modes inside yttrium iron garnet disk and ring microcavities of various sizes and found a significant improvement in modes coupling up to $\sim4.5$ kHz. 
Maximal theoretical conversion efficiency for small disks with radius $5~\upmu$m can reach unity for optimal optical power $\sim100~\upmu$W, which is experimentally feasible. Strategies for further improvements of interactions are discussed. 

\end{abstract}

\maketitle

\section{Introduction}
Modern quantum technologies open up opportunities for non-classical calculations that complement boolean logic~\cite{ladd2010quantum}. To build a complete quantum computer and quantum internet it is crucial to connect different computation devices~\cite{azuma}. Such a connection is expected to be at the speed of light which is conventionally achieved in optical fibers. To transmit information using optical networks a conversion of electromagnetic signal from quantum core with a typical frequencies up to several GHz~\cite{ladd2010quantum} to hundreds-THz region is needed. Such a conversion is implemented via quantum interconnects that bidirectionally convert quantum states from one physical system to another~\cite{kimble, lambert}. 

\par One of the leading platforms for quantum computation is superconducting one, on which quantum supremacy was recently demonstrated~\cite{Arute}. The main drawback of the superconducting platform is low operation temperature (less than 100 mK), which is governed by the microwave frequency of the qubit transition~\cite{krantz}. Microwave photons are hard to transmit over long distances due to large thermal population at room temperature which leads to thermal noise that destroys quantum information~\cite{lambert}. This obstacle limits realization of the quantum internet  using solely a superconducting platform.

\par At optical frequencies, thermal population is negligible, and quantum information doesn't suffer from thermal noise. Modern optical technologies allow the low-loss photon transmission ~\cite{schuster2014material, chen}, its detection at single particle level~\cite{hadfield} and quantum memory implementation~\cite{lvovsky}. For these reasons, solving the problem of coherent bidirectional conversion from microwave to optical range is in high demand because it can accelerate the development of quantum technologies.

\par Quantum converters can be implemented using various physical platforms, including non-linear electro-optical coupling~\cite{rueda}, $\Lambda$-systems and Rydberg atoms ~\cite{williamson}, optomechanical coupling ~\cite{andrews} and piezoelectric systems ~\cite{vainsencher}. Recently, coherent coupling between superconducting qubit and magnon was demonstrated~\cite{tabuchi,xu2023quantum,xu2024macroscopic}. This attracted attention to microwave-optical conversion using optomagnonic interaction~\cite{hisatomi}. The main feature of the optomagnonic coupling is flexibility in magnon frequency tuned by external magnetic field~\cite{gurevich2020magnetization}.

\par Optomagnonic coupling requires long-lived optical and magnon modes~\cite{rameshti2022cavity}. Ferromagnetic insulators, such as iron garnets, are among the most appropriate materials for this application.~\cite{gilleo1980ferromagnetic}. Yttrium iron garnet (YIG) is an ideal host for magnons due to its high spin density and extremely low Gilbert damping \cite{stancil}. The high spin density enables strong coupling of magnon modes with other excitations ~\cite{lachance}. Furthermore, because this material is optically transparent in the visible and near-infrared ranges, it is used in a variety of magneto-optical devices, from chemical and biosensors~\cite{ignatyeva2016magneto, ignatyeva2021vector} to on-chip optical isolators~\cite{bi2011chip, pintus2025integrated}.

\par The first optomagnonic coupling experiments were performed using an YIG sphere, yielding a conversion efficiency of $\eta=10^{-10}$, which describes the fraction of microwave photons that are converted to the optical range~\cite{hisatomi}. It was then proposed that using a high-quality optical mode in the form of a whispering gallery mode (WGM) would significantly increase the conversion rate. Later experiments showed no significant increase in coupling strength (single-photon coupling constant did not exceed 10 Hz and $\eta$ did not exceed $10^{-7}$)~\cite{osada2016,zhang2016,haigh2016}. Similar values were obtained also for a magnet-optical waveguide~\cite{zhu2022inverse} and for thin YIG film in an open cavity~\cite{haigh2020subpicoliter}. The main reason for such a weak interaction was a small overlap between optical and magnon modes. The magnon Kittel mode occupies the entire volume of the YIG sphere, whereas WGM occupies only a small part near the sphere's equator. A possible way to improve this situation is to use higher order magnetostatic modes that are concentrated near the sphere's boundary (such as Damon-Eshbach modes)~\cite{osada2018prl, graf2018cavity, sharma2019optimal}. According to ~\cite{osada2016}, using a ferromagnetic disk that supports WGM can increase conversion efficiency by up to $\eta=3\times10^{-2}$. Thin YIG disks have a much smaller volume compared to a sphere of the same radius, resulting in a higher coupling constant~\cite{wu2021optimal}. A compact on-chip optomagnonic transduction requires the fabrication of disk resonators on a substrate which is typically made of crystalline $\mathrm{Gd}_3\mathrm{Ga}_5\mathrm{O}_{12}$ (gadolinium gallium garnet, GGG).

\par In this study, we theoretically show possibility of photon-magnon interaction increase in disk and ring YIG microresonators. The calculation focused on high-quality films grown on a crystalline substrate, such as GGG, with the optical mode concentrated in the garnet layer. A thorough analysis of the magnon and optical modes overlap, and the modes volume in YIG microdisks results in single-photon coupling constants increasing up to several kHz.

\section{Results and discussion}
\subsection{Optomagnonic interaction}
\par The interaction between magnons and light is indirect. To accurately characterize the coupling rate, a general Hamiltonian of the electron system, electromagnetic radiation, and their interaction should be introduced. For the spin part of electronic system that interacts with external magnetic fields and near-neighbor the exchange coupling gives the Heisenberg Hamiltonian, which describes ferromagnets~\cite{Kittel}. In real samples with finite dimensions and shapes, the contribution of the magnetic dipole-dipole interaction and the energy of magnetic anisotropy should also be taken into account~\cite{gurevich2020magnetization}. Using the Holstein-Primakoff approximation \cite{HP}, a ferromagnet's Hamiltonian can be expressed in terms of bosonic creation $\hat{m}^{\dagger}_{\vec{k}}$ and annihilation operators $\hat{m}_{\vec{k}}$. A set of harmonic oscillators describes the Hamiltonian in its first order of expansion. The degeneracy of the modes is lifted for different momentum vectors $\vec{k}$ \cite{stancil}. In this paper, we will restrict ourselves to only the mode with $\vec{k} =0$, which is the homogeneous Kittel mode.

\par Optical part of the Hamiltonian can also be quantized and represented as a set of harmonic oscillators \cite{scully}. At optical frequencies the electric dipole interaction itself can change the number of magnons only via the spin-orbit interaction. The magnetic dipole interaction between the magnetic component of the optical field and the electrons subsystem of the ferromagnetic material is negligible~\cite{fleury1968scattering}. As a result, magnons have been found to be coupled with light through inelastic scattering described by the process within the second order of perturbation theory~\cite{shen1966interaction}. In the vast majority of cases, the light-magnon interaction can be described by the electric field displacement dependence on magnetization. This approach includes the dependence of the dielectric tensor on the magnetization~\cite{Legall}. Expanding the dielectric tensor in powers of $\vec{M}$ produces various magneto-optical effects, including Faraday, polar and longitudinal Kerr, and Cotton-Mouton effects caused by elastic light scattering and inelastic Brillouin light scattering (BLS) on magnons~\cite{wettling}. 
In this paper we will restrict ourselves to considering the contribution of only the first order of expansion to the optomagnonic coupling since the Cotton-Mouton contribution is typically small in iron garnets. 
For YIG, which is a cubic crystal, dielectric permittivity can be expressed as following~\cite{zvezdin1997modern}:
\begin{equation}
\label{eq1_perm_tens}
    \varepsilon_{ij}(\vec{M})=\varepsilon\delta_{ij}-if\sum_{k}\epsilon_{ijk}M_k.
\end{equation}
Here 
$\varepsilon$ is permittivity of the non-perturbed media, $\delta_{ij}$ is Kronecker delta, $f$ is a magneto-optical constant, $\epsilon_{ijk}$ is Levi-Civita symbol, $M_k$ is magnetization component. Magnon modes are considered to be small fluctuations ($\delta\vec{M}$) around the ground state ($\vec{M}_0$) \cite{graf2021design}:
\begin{equation}
\label{eq2_dyn_magnet}
    \vec{M}(\vec{r},t)=\vec{M}_0\sqrt{1-\left|\frac{\delta \vec{M}(\vec{r},t)}{M_s}\right|^2}+\delta\vec{M}(\vec{r},t).
\end{equation}
The ground state satisfies the equality $\vec{M}_0^2=M_s^2$. Magnetization is assumed to be parallel to z-axis, so that $\vec{M}_0=M_s\vec{e}_z$. Small fluctuations $\delta\vec{M}$ are perpendicular to the ground state. Thus, magnetization dependent part of the permittivity tensor can be written as follows: 
\begin{equation}
\label{eq3_optomagn_tensor}
\overleftrightarrow{\varepsilon}^{(1)}(\vec{M})=\begin{pmatrix}
0 & -ifM_s & if\delta M_y\\
ifM_s & 0 & -if\delta M_x\\-if\delta M_y&if\delta M_x&0
\end{pmatrix}.
\end{equation}

This term contributes to the energy of the whole system:  
\begin{equation}
    H=\frac{\varepsilon_0}{2}\int dV\vec{E}\overleftrightarrow{\varepsilon}^{(1)}\!(\vec{M})\vec{E},
    \label{eq4_Energy from M}
\end{equation}
where $\varepsilon_0$ is vacuum permittivity. Electric field can be quantized using a standard procedure~\cite{scully}:
\begin{equation}
\label{eq5_E_quant}    \hat{\vec{E}}(\vec{r},t)=i\sum_k\sqrt{\frac{\hbar\omega_k}{2V\varepsilon \varepsilon_0}} \left( \hat{a}_{k}\vec{u}_{k}(\vec{r})e^{-i\omega_kt} -\hat{a}^{\dagger}_{k}\vec{u}^{*}_{k}(\vec{r})e^{i\omega_kt} \right).
\end{equation}
In Eq.~\eqref{eq5_E_quant} $\hbar$ is the Plank constant, $\vec{u}_{k}(\vec{r})$ is normalized distribution of a $k$-th optical mode, $\omega_k$ is frequency of the optical mode, $V$ is volume occupied by the mode, $\hat{a}_{k}$ and $\hat{a}^{\dagger}_{k}$ are annihilation and creation operators for photons.
For dielectric resonators like YIG, it is convenient to use an effective mode volume defined as:
\begin{equation}
\label{eq6_opt_mode_vol}
V_k=\frac{\int\left|\vec{E}_{k}(\vec{r})\right|^2 dV}{\text{max}\left|\vec{E}_{k}(\vec{r})\right|^2}.
\end{equation}
If the energy stored in a single mode is the energy of a single photon, we obtain the corresponding amplitude $\vec{E}_k(\vec{r})=\sqrt{\frac{\hbar\omega_k}{2V\varepsilon \varepsilon_0}}\vec{u}_{k}(\vec{r})$ and normalization condition for overlap functions: \cite{safavi2014optomechanical}:
\begin{equation}
\label{eq7_opt_norm}
    \int \vec{u}_{k}(\vec{r})\vec{u}^*_{k'}(\vec{r})dV=\delta_{kk'}V_k.
\end{equation}
For small fluctuation $|\delta\vec{M}|\ll M_s$ the quantization of magnetization fluctuation takes the following form:
\begin{equation}
    \delta\vec{M}(\vec{r},t)=M_s\sum_{l}\left(\delta \vec{m}_{l}(\vec{r})\hat{b}_l e^{-i\omega_l t}+\delta \vec{m}^{*}_{l}(\vec{r})\hat{b}^{\dagger}_l e^{+i\omega_l t}\right).
\label{eq8_M_quant}
\end{equation}

This leads to the magnons Hamiltonian $H_m=\hbar\sum_l\omega_l \hat{b}^\dagger_l \hat{b}_l$ and normalization condition:
\begin{equation}
\label{eq9_magnon_norm}
    iM_s\int\vec{m}_0[\delta \vec{m}_{\gamma}(\vec{r})\times \delta \vec{m}_{l}^{*}(\vec{r})] dV=g\mu_B\delta_{\gamma l},
\end{equation}
where $\vec{m}_0=\vec{M}_0/M_s$, $g $ is the Landé factor and $\mu_B$  is the Bohr magneton.
After the substitution of equations ~\eqref{eq3_optomagn_tensor},~\eqref{eq5_E_quant},~\eqref{eq8_M_quant} into~\eqref{eq4_Energy from M}  two contributions depending on the number of bosonic operators can be separated. The first comes from the tensor components which are proportional to $M_s$:
\begin{multline}
\label{eq10_log_Ham}
    \hat{H}^{(I)}=-\sum_{k,k'}\frac{ifM_s\hbar\sqrt{\omega_k\omega_{k'}}}{4\varepsilon\sqrt{V_kV_{k'}}}\int dV\\ \Big(\hat{a}_{k}^{\dagger}\hat{a}_{k'} \left[\vec{u}_{k}^{*}(\vec{r})\times \vec{u}_{k'}(\vec{r}) \right]_z e^{i(\omega_k-\omega_{k'})t}+\\
    +\hat{a}_{k}\hat{a}_{k'}^{\dagger} \left[\vec{u}_{k}(\vec{r})\times \vec{u}_{k'}^{*}(\vec{r}) \right]_z e^{-i(\omega_k-\omega_{k'})t}+\\+\hat{a}_{k}\hat{a}_{k'} \left[\vec{u}_{k'}(\vec{r})\times \vec{u}_{k}(\vec{r}) \right]_z e^{-i(\omega_k+\omega_{k'})t}+\\+\hat{a}_{k}^{\dagger}\hat{a}_{k'}^{\dagger} \left[\vec{u}_{k'}^{*}(\vec{r})\times \vec{u}_{k}^{*}(\vec{r}) \right]_z e^{i(\omega_k+\omega_{k'})t}\Big).
\end{multline}
Considering a time interval much longer than the period of fast oscillations with frequency $\omega_k+\omega_{k'}$ but much shorter than the period of slow oscillations $\omega_k-\omega_{k'}$ and making time-averaging, the third and fourth terms in  ~\eqref{eq10_log_Ham} can be shorten to the following form:
\begin{equation}    \hat{H}^{(I)}=\hbar\sum_{k,k'}g_{k,k'}\hat{a}_{k}^{\dagger}\hat{a}_{k'} +h.c.,
\end{equation}
where $g_{k,k'}=\Omega \int dV \left[\vec{u}_{k}^{*}(\vec{r})\times \vec{u}_{k'}(\vec{r}) \right]_z$ and  $\Omega=-\frac{ifM_s\sqrt{\omega_k\omega_{k'}}}{4\varepsilon\sqrt{V_kV_{k'}}}$.
This part of the Hamiltonian describes the static Faraday rotation. Another part involving small fluctuations of magnetization describes inelastic magnon-photon scattering, i.e. BLS. After making time-averaging similar to that for Hamiltonian $H^{(1)}$, we obtain:
\begin{equation}
    \hat{H}^{(II)}=-\hbar\sum_{k,k',l}\Big(G_{k,k',l}^{+}\hat{a}_{k}^{\dagger}\hat{a}_{k'}\hat{b}_l+G_{k,k',l}^{-}\hat{a}_{k}^{\dagger}\hat{a}_{k'}\hat{b}_l^{\dagger}\Big)+h.c.,
\end{equation}
where $G_{k,k',l}^{+}=\Omega\int \left[\vec{u}_{k}^{*}(\vec{r})\times \vec{u}_{k'}(\vec{r}) \right]\delta\vec{m}_l (\vec{r})dV$ and  $G_{k,k',l}^{-}=\Omega\int \left[\vec{u}_{k}^{*}(\vec{r})\times \vec{u}_{k'}(\vec{r}) \right]\delta\vec{m}_l^{*} (\vec{r})dV$. 
\par YIG, being a dielectric, can serve as a resonator for optical photons participating in BLS and thereby lead to an increase in the optomagnonic coupling.  It is critically important that both optical and magnon modes should satisfy the energy conservation law and selection rules~\cite{osada2018njp}. The former one is known as the triple resonance condition, which for the Stokes and anti-Stokes process has form $\omega_k=\omega_{k’}\pm\omega_l $~\cite{haigh2016}. 
We can express for circularly polarized magnons $\delta \vec{m}_l=1/\sqrt{2}(\vec {e}_x+i\vec {e}_y)\delta m =\vec {v}_l\delta m$, where $\delta m=\sqrt{\frac{g\mu_B}{VM_s}} $ is obtained from the normalization condition. Introducing  $V_m=\int |\vec{v}_l (\vec{r})|^2 dV $ and interaction volume: 
\begin{equation}
    V_{int}=\int \left[\vec{u}_{k}^{*}(\vec{r})\times \vec{u}_{k'}(\vec{r}) \right]\vec{v}_l (\vec{r})dV.
    \label{interaction volume}
\end{equation}
Optomagnonic coupling, for example, for the scattering process of a photon in some mode $k'$ into a photon in mode $k$ via a magnon in mode $l$ that satisfies selection rules (conservation of total angular momentum) and the triple-resonance condition, can be written in the following way:
\begin{multline}
      G_{k,k',l}^{+}=-\frac{ifM_s}{4\hbar }\sqrt{\frac{4g\mu_B}{M_sV_m}}\sqrt{\frac{\hbar\omega_k}{2\varepsilon V_k}}\sqrt{\frac{\hbar\omega_{k'}}{2\varepsilon V_{k'}}} V_{int}.
      \label{coupling}
\end{multline}

As can be seen from the equations \eqref{interaction volume} and \eqref{coupling}, the optomagnonic coupling value depends on the modes overlap and their volumes. To improve coupling, one should increase the overlap and reduce the volume of modes.

\subsection{Whispering gallery modes in YIG disks/rings}
\par Since the 1960s, high-quality dielectric resonators shaped like disks with whispering gallery modes (WGM) have gained significant attention in the microwave frequency range ~\cite{cohn1968microwave}. The first successful implementation for the optical frequency range occurred in the beginning of the 2000s ~\cite{borselli2004rayleigh}. An exact analytical solution for the eigenmodes of a finite dielectric microdisk and microrings is impossible. 
Therefore, to find the WGM frequencies and field distribution, numerical methods are usually used, that are based on either finite element method (FEM) or finite-difference time-domain (FDTD) approach. 

\begin{figure*}[ht]
\centering
\includegraphics[width=0.8\textwidth]{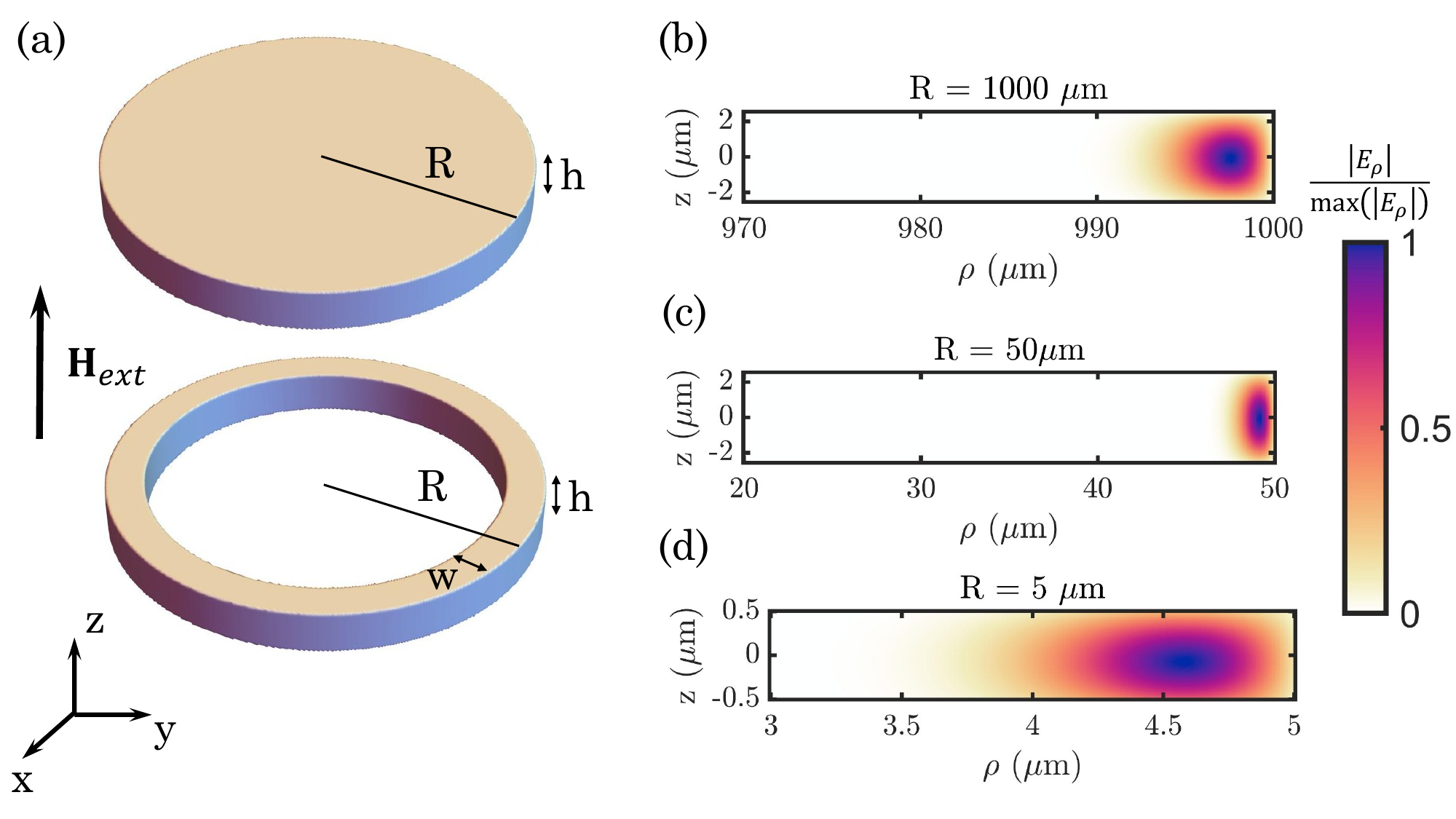}
\caption{(a) Disk and ring geometry of the  resonators under consideration. (b-d) Characteristic normalized distribution of the TE WGM mode $E_{\rho}$ component for the YIG disks on GGG (b,c) and ring (d). For all distributions $p = 1, q = 1$ ($m$ varies and depends on the radius and height of the resonator). All distributions are plotted in $z-\rho$ cross-section.}
\label{fig1: structure and modes}
\end{figure*}

\par To calculate the magnon-photon coupling constants, WGM modes were simulated for YIG disk and ring resonators with different diameters using the FEM method. For this purpose, an eigenmode analysis based on the wave equation in a system owing axial symmetry was performed:
\begin{equation}
\label{eq15_wave_eq}   \vec{\nabla}\times\vec{\nabla}\times\vec{E}-k_0^2\overleftrightarrow{\varepsilon}\vec{E}=0.
\end{equation}
The solution of \eqref{eq15_wave_eq} in an axisymmetric system is a standing wave $\vec{E}(\vec{r}) = \vec{E}(\rho,z)e^{-im\phi}$, where $m$ is the azimuthal mode number. $\vec{E}(\rho,z)$ itself depends on the polar ($p$) and radial ($q$) mode number (see Appendix A). In our calculation we are focused on the fundamental TM and TE modes with wavelength near 1.55 $~\upmu$m (C-band of optical communications) which possess the highest quality factor~\cite{gorodetsky2011optical}, so $p=q=1$. TE and TM modes of the resonator possess $E_{\phi}$, $E_z$, $H_{\rho}$,  and $E_{\rho}$, $H_{\phi}$, $H_z$ respectively. The modes are eigenmodes of the resonator, and their frequency difference can be matched to the magnon frequency (i.e., meet the triple-resonant condition) in the BLS process by a combination of free spectral range and geometrical birefringence. Moreover, since these modes are orthogonal, they give a non-zero contribution to the overlap integral. Numerically calculated eigenfrequencies were verified with the ones obtained by a semi-analytical approach (see Appendix A).

\par Permittivity of the YIG was set to be constant and equal to $\varepsilon=4.84$ which is valid for wavelength $\sim1.5$ $\upmu$m ~\cite{wemple1974optical}. The radius of the resonators varied from 5 to 1000 $\upmu$m. To satisfy the condition $R >> h$ the thickness of both systems $h$ was set to 5 $\upmu$m for $50\leq R\leq1000$ $\upmu$m ("large" disks) and 1 $\upmu$m for $5\leq R\leq50$ $\upmu$m ("small" disks). The width of the ring resonator was set to 10 $\upmu$m for $50\leq R\leq1000$  and 2 $\upmu$m for the remaining cases. The thickness of the rings was chosen to maintain the homogeneity of the Kittel mode magnetization distribution~\cite{zhou2017spin, zhou2021engineering}. In this study, resonators with a radius less than 5 $\upmu$m were not considered because WGM can degenerate into Mie modes, which requires a special analysis~\cite{bohren2008absorption}. Disk resonators were simulated in free space and also supported by GGG substrates ($\varepsilon=3.88$). The first case physically corresponds to a resonator on a pedestal ~\cite{johnson2006self} while the latter one is its on-chip counterpart ~\cite{soltani2007ultra}. Ring resonators which are commonly used in modern integrated optics ~\cite{bogaerts2012silicon} were supported by GGG substrate. A schematic representation of all geometries is presented in Fig.~\ref{fig1: structure and modes}a.

\begin{figure*}[ht]
\centering
\includegraphics[width=1\textwidth]{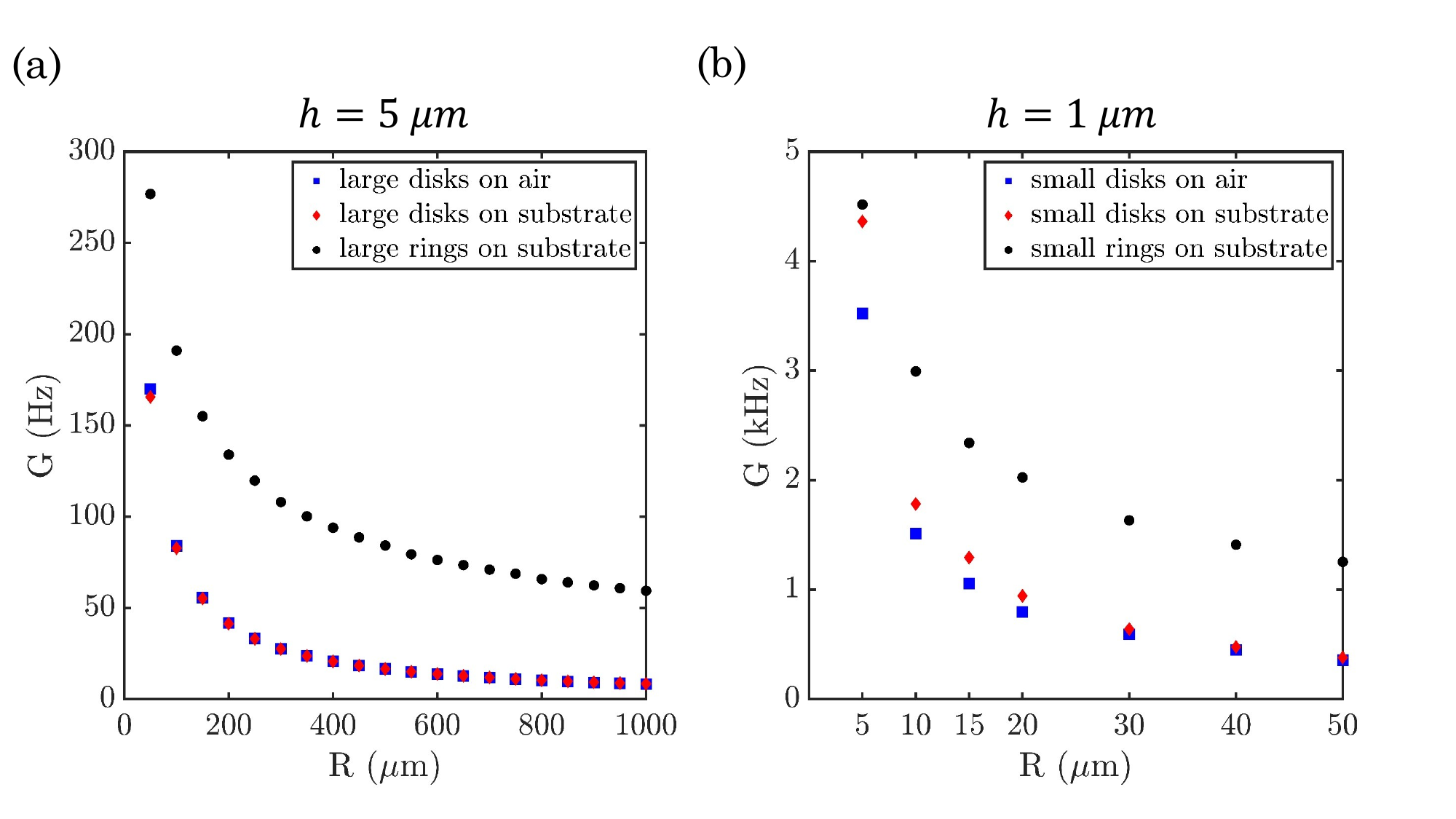} 
\caption{Coupling constant for large (a) and small (b) disks and rings.}
\label{fig3: coupling constant}
\end{figure*}

The WGM electromagnetic field distribution is distinguished by a high degree of localization near the disk corner as shown in Fig.~\ref{fig1: structure and modes}b-d. As the size of the disk/ring decreases, the electromagnetic field becomes more localized near the air-garnet boundary, leading to both physical and mode volume decrease (see Appendix B, Fig.~\ref{fig2: mode volume}). For smallest disk with $R=5~\upmu$m the mode volume drops down to $0.01~\upmu \text{m}^3$, which is several orders lower than the one obtained in large YIG spheres and also close to the values achieved in the open resonator containing a thin YIG film on a substrate located between two Bragg mirrors~\cite{haigh2020subpicoliter}. The Kittel magnon mode occupies the entire volume of the resonator, which means that switching to disk geometry decreases magnon mode volume by a factor of $\frac{V^{sphere}_{m}}{V^{disk}_{m}}=\frac{4R}{3h}$. For the large disks with $h = 5~\upmu$m this relation is larger than 250. For the small disks with $1~\upmu$m thickness the $\frac{V^{sphere}_{m}}{V^{disk}_{m}}$ maximum value is close to 7. A similar situation is observed for the rings. Consequently, planar disk and ring geometries can give substantial coupling constant increase in terms of magnon mode volume. However,  because the electromagnetic field of the WGM mode is localized near the YIG-air interface, transition from disk to ring geometry has no substantial effect on the optical mode volume.

\subsection{Coupling constants and conversion efficiency}

\par  To calculate the coupling constant $G$ it is convenient to pass from $\vec{E}_k(\rho, \phi, z)$ to dimensionless $\vec{u}_k(\rho, \phi, z)$ which was performed using procedure mentioned in section II.A. Here $N$ is the photon number that can be obtained from the energy of the electric field inside the cavity $\frac{\varepsilon_0\varepsilon}{2}\int |E_k|^2dV = \frac{1}{2}\hbar\omega_k N$. As the system possesses axial symmetry, a uniform precession of a magnon Kittel mode with right or left circular polarization in the plane of the disk/ring takes the following form:
\begin{equation}
\label{eq16_magnon_in_cyl_coord}
    \vec{v}_l(\rho, \phi, z) = \frac{1}{\sqrt{2}}e^{\pm i\phi}\vec{e}_\rho \pm \frac{i}{\sqrt{2}}e^{\pm i\phi}\vec{e}_\phi.    
\end{equation}
Applying the magnon mode polarization vector to Eq. \eqref{interaction volume} yields an angular moment selection rule for the Stokes and anti-Stokes processes. For the anti-Stokes process a scattering of the photons of an optical mode with clockwise direction of propagation on the left-handed polarized magnon mode results in a non-zero $G^+$ constant while for the right-handed magnons $G^+$ becomes zero. Similarly for the opposite polarization. Optomagnonic coupling constant $G$ significantly depends on the overlap integral which for the case of a WGM resonator supporting $TE_{m+1}$ $(u_{k})$ and $TM_m$ $(u_{k'})$ optical modes simplifies to the following form:
\begin{multline}
\label{eq17_Vint_WGM}
    V_{int} = \sqrt{2}\pi\int\int \Big(\Tilde{u}_{k\varphi}^* \Tilde{u}_{k'z}+i  \Tilde{u}_{k\rho}^*\Tilde{u}_{k'z})\Big)\rho dh d\rho,
\end{multline}
where $u_{kj}=\Tilde{u}_{ki} e^{-i m \phi}$, where $j=\{\rho,\phi,z\}$.
\par Fig.~\ref{fig3: coupling constant} shows the coupling constant as a function of the resonator radius for the disk and ring geometries. The coupling constant values for disks with and without substrates do not differ significantly. The coupling constant for disks with a radius comparable to the YIG spheres ($R\sim 150-300 ~\upmu$m) is more than 5 times larger than for the YIG spheres: $27-55$ Hz vs $\sim 5-10$ Hz, respectively \cite{osada2016, zhang2016, wu2021optimal}). Ring geometry reduces physical volume of the resonator. This results in a decrease in the Kittel mode volume ($V_m = V_{res}$) while $V_{int}$ remains close to the disk geometry. Consequently, for $R\sim 150-300 ~\upmu$m the ring geometry provides additional coupling constant improvement up to $\sim5$ times. The situation changes dramatically for a smaller radius (Fig.\ref{fig3: coupling constant}b). For such a small resonators $G$ tends to reach kHz values ($\sim 4.5$ kHz for radius $5 ~\upmu$m disk/ring). Notably, there is no significant difference between the $G$ value for the ring and the disk for such small resonators. The main reason for this is the large width of the ring ($2~\upmu$m) which brings the magnon mode volumes close together. Potentially, the width of the ring can also be submicron. However, in such a case, a thorough analysis of the magnon modes should be carried out taking into account exchange interaction which is responsible for the standing spin waves~\cite{zhou2021engineering}.

\par The conversion efficiency of a quantum optomagnonic transducer is the ratio of output optical photons to input microwave photons following a transduction process. It defines the efficiency of converting one type of quantum signal to another while retaining quantum information. Magnon based microwave-optical conversion is a one-stage process, so conversion efficiency can be written as follows ~\cite{han2021microwave}:
\begin{equation}
     \eta = \eta_o \eta_e \frac{4C_{om}C_{em}}{(1+C_{om}+C_{em})^2}
     \label{conversion efficiency}.
\end{equation}
Here, $\eta_o$ and $\eta_e$ are the external coupling rates for optical and microwave modes, respectively. $C_{e(o)m}=\frac{4G^2_{e(o)m, MP}}{\kappa_{e(o)} \kappa_m}$ is the cooperativity between photonic (e - microwave, o - optical) and magnonic $m$ modes. $G_{e(o)m, MP} = G_{e(o)m}\sqrt{N}$ is multiphoton coupling constant, $\kappa_{e(o)}$ and $\kappa_m$ are dissipation rates of the corresponding modes. Here we consider only interaction between optical photons and magnons, so Eq.~\ref{conversion efficiency} can be reduced to the following:
\begin{equation}
    \eta = \eta_o  \frac{4C_{om}}{(1+C_{om})^2}
    \label{optomagnonic_conversion}.
\end{equation}
To reach unity conversion efficiency for the fixed single-photon coupling constant equation $\frac{d\eta}{dN}=0$ can be solved. The solution gives $N=\frac{\kappa_o\kappa_m}{4G^2}$ as an optimal value for fixed $\kappa_o$, $\kappa_m$ and $G$.

\begin{figure}[t]
\centering
\includegraphics[width=0.4\textwidth]{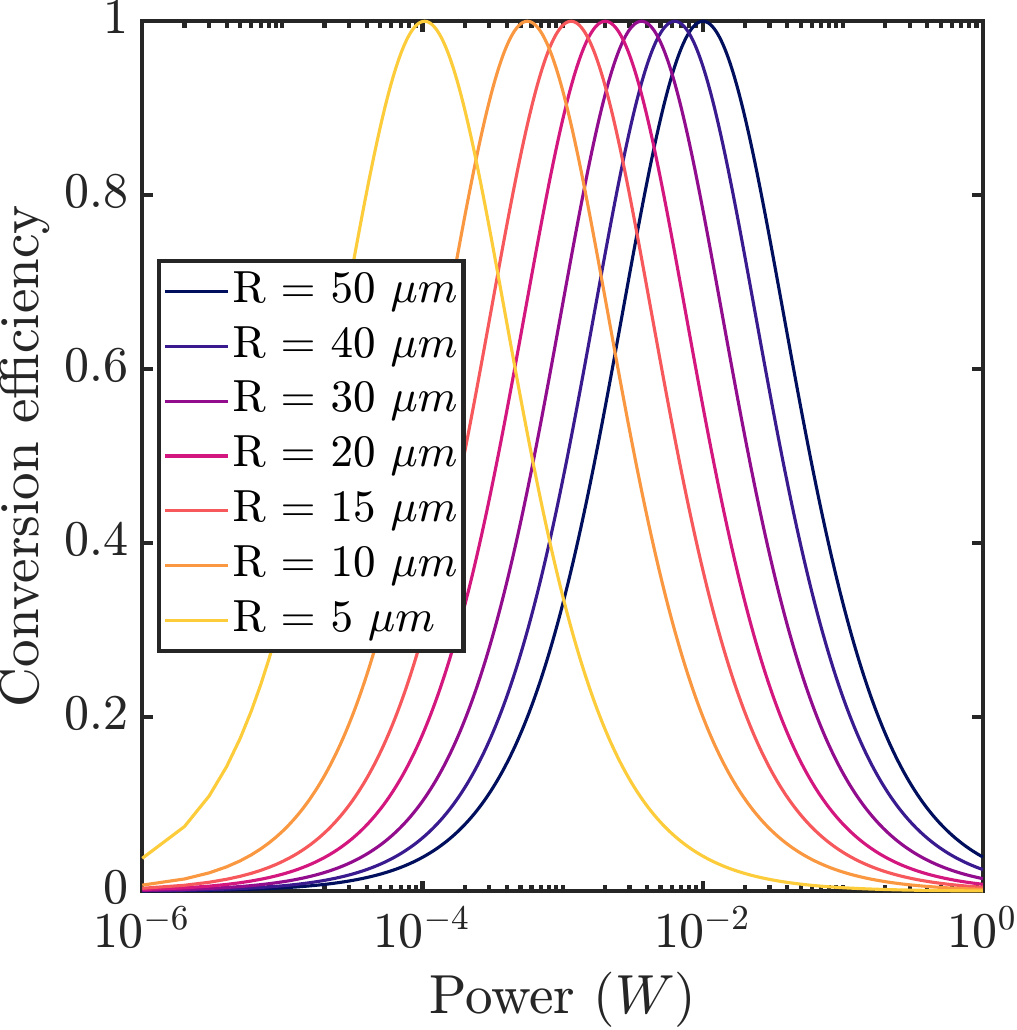} 
\caption{Conversion efficiency as a function of optical power inside the resonator for small disks with different radius.}
\label{fig3: converstion_efficiency}
\end{figure}

The dissipation rate of an optical WGM is affected by various parameters, including material absorption, mode leakage, and surface inhomogeneities that cause mode scattering~\cite{gorodetsky2011optical}. Proper surface treatment of a spherical resonator reduces surface scattering effects, therefore only internal losses are responsible for the mode dissipation~\cite{zhang2016, rameshti2022cavity}. In the case of a disk or ring resonator the electron and ion beam lithography techniques allow fabrication of high-quality structures with roughness typically less than 1 nm~\cite{ignatyeva2020all}. Consequently, only internal losses caused by photon absorption can be considered. The WGM dissipation rate can be expressed using the resonator quality factor $\kappa_o=\frac{\omega_0}{Q_{o, int}}=\frac{c\alpha}{2\pi n}$. The absorption coefficient $\alpha$ of YIG at $1.55~\upmu$m is approximately $0.1~\text{cm}^{-1}$~\cite{rameshti2022cavity, doormann1984measurement, wood1967effect} which leads to $\kappa_o\sim 0.2$ GHz at this wavelength. Magnon internal dissipation normally does not exceed several MHz, and even at low temperatures, it can approach 1 MHz~\cite{kosen2019microwave}. Fig.~\ref{fig3: converstion_efficiency} shows conversion efficiency as a function of optical power inside the small disk resonators. For each resonator size, there is an optimal value of the optical power. The smaller the resonator, the less power is needed to achieve unity conversion. For the smallest one with radius $R=5~\upmu$m, optical power $P\sim100~\upmu$W is enough to reach $\eta=1$. In that case, it gives the multiphoton coupling constant that is estimated to be $G_{om, MP}=G^+\sqrt{N}=G^+\sqrt{\frac{P}{\kappa_o \hbar \omega_0}} = 8.49~\text{MHz}$. For large resonators the optimal power significantly exceeds the $\text{mW}$ range, making them unsuitable for working in the quantum regime. For large resonators the optimal power significantly exceeds the $\text{mW}$ range, making them unsuitable for working in the quantum regime. The increased power leads to non-linear absorption and the Kerr effect, which affects the resonance line shape and reduces the optical quality factor of the microresonator \cite{grudinin2006ultrahigh}. Moreover, the thermo-optic effect can lead to a significant frequency drift \cite{kippenberg2004nonlinear} of the resonance, which may ultimately lead to a breach of the triple resonance condition.

\section{Conclusion}
In this study we demonstrated by numerical simulation that optomagnonic coupling in YIG resonators supporting optical whispering gallery modes can be significantly enhanced through transition from spherical to disk and ring geometries. The YIG sphere resonators with typical size less then $300~\upmu$m are difficult to produce and handle. On the contrary, their on-substrate disk or ring counterparts can be manufactured using standard electron, ion beam, and photolithography techniques ~\cite{khramova2022accumulation, ignatyeva2020all, bi2011chip} and incorporated into integrated optical systems, making them appealing for device applications. Moreover, for disk of radius similar to available YIG spheres the improvement is more than fivefold. A switching to ring geometry additionally enhances this value by five times. In the ultimate case of the small thin disks/rings with radius of about 5 $\upmu$m the coupling constant reaches $\sim 4.5~\text{kHz}$. For a quite moderate optical power of $\sim100~\upmu$W inside the resonator conversion efficiency is estimated to reach unity.
\par Several strategies can be used for further optomagnonic coupling enhancement. The first approach involves utilization of whispering gallery magnon modes which were recently observed in permalloy ($\mathrm{Ni}_{81}\mathrm{Fe}_{19}$) microdisks ~\cite{schultheiss2019excitation}. Another way is to use optical Mie modes in nanoparticles. These resonators possesses extremely small optical and magnon mode volumes ~\cite{krichevsky2024inverse, ignatyeva2024optical, almpanis2020spherical}.

\section*{acknowledgement}

The authors thank A. Shitikov, A. Vorobyev, D. Ignatyeva and A. Kalish for fruitful discussion on whispering gallery modes resonators physics. The work is financially supported by the Russian Science Foundation, project No. 25-22-00253. 

\appendix
\section{Analytical approximation for the WGM modes eigenfrequency calculation}
The wave equations for optical cavities in the general case can be written in the form:
\begin{align}
         & \vec{\nabla}\times\vec{\nabla}\times\vec{E}-k_0^2\overleftrightarrow{\varepsilon}\vec{E}=0, \label{eq16}\\ 
     & \vec{\nabla}(\overleftrightarrow{\varepsilon}\vec{E})=0,\\
     & \vec{\nabla}\times(\overleftrightarrow{\varepsilon}^{-1}\vec{\nabla}\times\vec{H})=0,\\
     & \vec{\nabla}\vec{H}=0.
\end{align}
For the microdisk resonator, it's convenient to use a cylindrical system of coordinates in which the solution of the Helmholtz vector equation is based on a scalar Helmholtz equation that, in particular, satisfies the $z$ component of the electric field. Particular solution of scalar equation has form:
\begin{equation}
    \psi(\rho,\phi,z)=Z_m(\sqrt{k^2-\beta^2}\rho)e^{-im\phi+i\beta z},
    \label{scalar potential}
\end{equation}
where $k^2=k_0^2\varepsilon$, $\beta$ is the propagation constant along the $z$-axis, $m$ is an integer number, which we will see has the meaning of azimuthal number, and $Z_m$ is the solution of the Bessel equation. The choice of a particular type of Bessel function depends on the boundary conditions and the required behavior at zero and infinity.
The general solution for the electric field can be expressed via scalar potential as:
\begin{align}
    &\vec{E}=C_{TE}\vec{M}+C_{TM}\vec{N}, \label{Debye potential1}\\
    &\vec{M}=\vec{\nabla}\times(\vec{e}_z \psi) \label{Debye potential2},\\
    &\vec{N}=\frac{1}{k}\vec{\nabla}\times\vec{\nabla}\times(\vec{e}_z \psi),  \label{Debye potential3} 
\end{align}
If the electric field $\vec{E}$ is expressed solely by $\vec{M}$, so it doesn't have an $E_z$ component and is called transverse electric (TE), if expressed solely by $\vec{N}$, it is called transverse magnetic (TM). Usually, WGMs in axially symmetric bodies are hybrids of TE and TM, but one of the energy integrals, $\frac{\varepsilon_0\varepsilon}{2}\int E_z^2dV$, $\frac{\mu_0\mu}{2}\int H_z^2dV$ is much greater than the others, so we can say that that mode is close to the TE or TM type.

Using equations \eqref{scalar potential}-\eqref{Debye potential3}, we can write down expressions for TE and TM modes in an infinite cylinder:
\begin{align}
\vec{E}_{TE}&=C_{TE}\left(\frac{im}{\rho}Z_m(k_{\rho}\rho)\vec{e}_{\rho}-\frac{\partial Z_m(k_\rho\rho)}{\partial\rho}\vec{e}_{\phi}\right)e^{-im\phi+i\beta z},\\
\begin{split}\label{eq:1}
\vec{B}_{TE}&=-\frac{i}{k_0c}C_{TE}\Big(i\beta\frac{\partial Z_m(k_\rho\rho)}{\partial\rho}\vec{e}_{\rho}-\frac{m\beta}{\rho}Z_m(k_{\rho}\rho)\vec{e}_{\phi}\\ &k_\rho^2Z_m(k_{\rho}\rho)\vec{e}_z\Big)e^{-im\phi+i\beta z},
\end{split}\\
\begin{split}\label{eq:3}
\vec{E}_{TM}&=-C_{TM}\frac{1}{\varepsilon\mu k_0}\Big(i\beta\frac{\partial Z_m(k_\rho\rho)}{\partial\rho}\vec{e}_{\rho}-\frac{m\beta}{\rho}Z_m(k_{\rho}\rho)\vec{e}_{\phi}\\ &k_\rho^2Z_m(k_{\rho}\rho)\vec{e}_z\Big)e^{-im\phi+i\beta z},
\end{split}\\
\vec{B}_{TM}&=C_{TM}\frac{i}{c}\left(\frac{im}{\rho}Z_m(k_{\rho}\rho)\vec{e}_{\rho}-\frac{\partial Z_m(k_\rho\rho)}{\partial\rho}\vec{e}_{\phi}\right)e^{-im\phi+i\beta z},
\end{align}
where $k_{\rho}=\sqrt{k^2-\beta^2}$, $C_{TE/TM}$ are normalization constants. The propagation constant $\beta$ can be real or imaginary depending on the boundary conditions. We are mostly interested in modes with large $m$ and with small $\beta$; in that case argument will stay real in all regions, and for describing fields outside the resonator, we will use the Hankel function.

In a finite disk, $\beta\neq0$, and modes are hybrid $\vec{E}=\vec{E}_{TE}+\vec{E}_{TM}$. To find a solution, we need to choose the proper form of Bessel function inside and outside of the disk and make them satisfy corresponding boundary conditions. The disk center is placed on the origin of the coordinate system and has a radius of $R$ and a height of $h$ and have the refractive index $n_1=n$, surrounded with air $n_0=1$. On the cylinder side walls there should be continuous tangential components of $E$ and $B$ and a normal component for $D$, which will bring us to the characteristic equation:
\begin{multline}
    \left(\frac{\frac{\partial J_m(y)}{\partial y}}{J_m(y)}-\frac{y}{x}\frac{\frac{\partial H_m^{(1)}(x)}{\partial x}}{H_m^{(1)}(x)}\right)\left(\frac{\frac{\partial J_m(y)}{\partial y}}{J_m(y)}-\frac{y}{n^2x}\frac{\frac{\partial H_m^{(1)}(x)}{\partial x}}{H_m^{(1)}(x)}\right)=\\=\frac{m^2}{n^2y^2}\left(1-\frac{y^2}{x^2}\right)\left(n^2-\frac{y^2}{x^2}\right),
\label{char eq 12}
\end{multline}
where  $y=k_{1\rho}R$, $x=k_{2\rho}R$. If $\beta\ll k$ second bracket on right side of equation \eqref{char eq 12} will approach to zero and which leads to the equation   splitting into two new equations, namely, the equality to zero of the first and second brackets on the left side of the equation \eqref{char eq 12}, which corresponds to the characteristic equations for the TE and TM modes. The WGM described with Bessel function with large azimutal number $m$ and resonance in the case $\beta<< k$ occurs near the first maximum of the Bessel function $y\approx m$. That means that $x<m$ and the Bessel and Neumann functions in that range of $x$ are not oscillating and can be approximated as \cite{olver2010nist}:
\begin{align}
    J_m(x)\approx\frac{1}{\sqrt{2\pi m}}\left(\frac{ex}{2m}\right)^m,\\
    Y_m(x)\approx-\frac{2}{\sqrt{2\pi m}}\left(\frac{ex}{2m}\right)^{-m}.
\end{align}
For large $m$ and $x<m$ the Bessel functions will be exponentially small and the Neumann functions in contrast are very large. That's why for finding real solutions we can approximate the Hankel function via the Neumann function and we arrive to characteristic equation for real eigenvalues:
\begin{equation}
    \frac{J_m^{'}(y)}{PyJ_m(y)}=\frac{Y_m^{'}(x)}{xY_m(x)},
\end{equation}
where $P=1$ for TE modes and $P=1/n^2$ for TM modes. For further proceed we need to use expression for derivative of Bessel function:
\begin{equation}
    J_m^{'}(y)=-\frac{m}{y}J_m(y)+J_{m-1}(y)
\end{equation}
and arrive to:
\begin{equation}
     \frac{J_{m-1}(y)}{J_m(y)}=\frac{m}{y}+\frac{PyY_m^{'}(x)}{xY_m(x)}.
     \label{eq33}
\end{equation}
On the right side of \eqref{eq33} in the region of resonance $x<m$ there is slowly varying function and we can use the Debye expansion \cite{olver2010nist} and obtain:
\begin{equation}
     \frac{J_{m-1}(y)}{J_m(y)}=\frac{m}{y}-\frac{Py}{x}\left(\frac{\sqrt{m^2-x^2}}{x}-\frac{x}{2(m^2-x^2)}\right).
     \label{eq34}
\end{equation}
For further analyze we need to make an assumption about large value of the refractive index $n\rightarrow \infty$ which bring us that for the TE modes the right side of \eqref{eq34} will be large and we seek solution for the TE mode near zeros of the Bessel functions $T_{m,q}$. For TM modes, in contrast, it will be small and allow us seek solution near $T_{m-1,q}$. Inserting this zero-order we can find first order correction to this solution and for large $m$ and not so large $q$ we can write solution for the TE and TM modes on same foot:
\begin{align}
y_{TE(TM)}=T_{m,q}-\Delta_{TE(TM)},\\
\Delta_{TE(TM)}=\frac{1}{P_{TE(TM)}n\sqrt{n^2-1}}.
\end{align}
In this approximation we can consider field and eigen frequencies equivalent  as in a closed resonator with increased radius by $\Delta_{TE,(TM)}\lambda/2\pi n$ with boundary conditions $H_n=E_n=0$.

Same procedure should be done on the cylinder bases for finding $\beta$. Since on that boundary we pass to $\rho=0$  so we should choose the Bessel function $J_m$ for describing fields. For satisfy continuity of $\vec{E}$ we need to take $k_{1\rho}=k_{3\rho}$ or $n^2k_0^2-\beta^2=k_0^2-\beta_3^2$ and write down proper boundary condition and obtain characteristic equation:
\begin{align}
    \sqrt{\frac{k_0^2(n^2-1)}{\beta_{TE}^2}-1}=\begin{matrix}
        \tan{(\beta_{TE}h/2)}\\ -\cot{(\beta_{TE}h/2}),
    \end{matrix}\\
    n^2\sqrt{\frac{k_0^2(n^2-1)}{\beta_{TM}^2}-1}=\begin{matrix}
        \tan{(\beta_{TM}h/2)}\\ -\cot{(\beta_{TM}h/2}).
    \end{matrix}
\end{align}
Graphic solution will give a hint about seeking solutions near $\beta h=(p+1)\pi$ and in the first-order approximation the dielectric cylinder is equivalent to a closed cavity with enhanced thickness $\Delta_{TM,(TE)}\lambda/2\pi n$ from two sides for the TE and TM modes, respectively. One can notice that indexes here are opposite with respect to the side walls of the cylinder. It is because the side walls and bases "see" opposite polarizations of the modes. This allows  getting a simple system for estimate frequencies and wavelengths $\lambda_{mqp}=2\pi n/k_{mqp}$ for the TE and TM modes by choosing corresponding $P_{TE(TM)}$ and solving system:
\begin{align}
&\beta_{mqp}=\frac{(p+1)\pi}{h+\frac{2P_{TE(TM)}n}{k_{mqp}\sqrt{n^2-1}}},\\
    &k_{mqp}R=\sqrt{\left(T_{mq}-\frac{1}{2P_{TE(TM)}n\sqrt{n^2-1}} \right)^2+\beta_{mqp}^2R^2}.
\end{align}

\section{WGM mode volume as a function of disk radius}

\begin{figure*}[ht]
\centering
\includegraphics[width=1\textwidth]{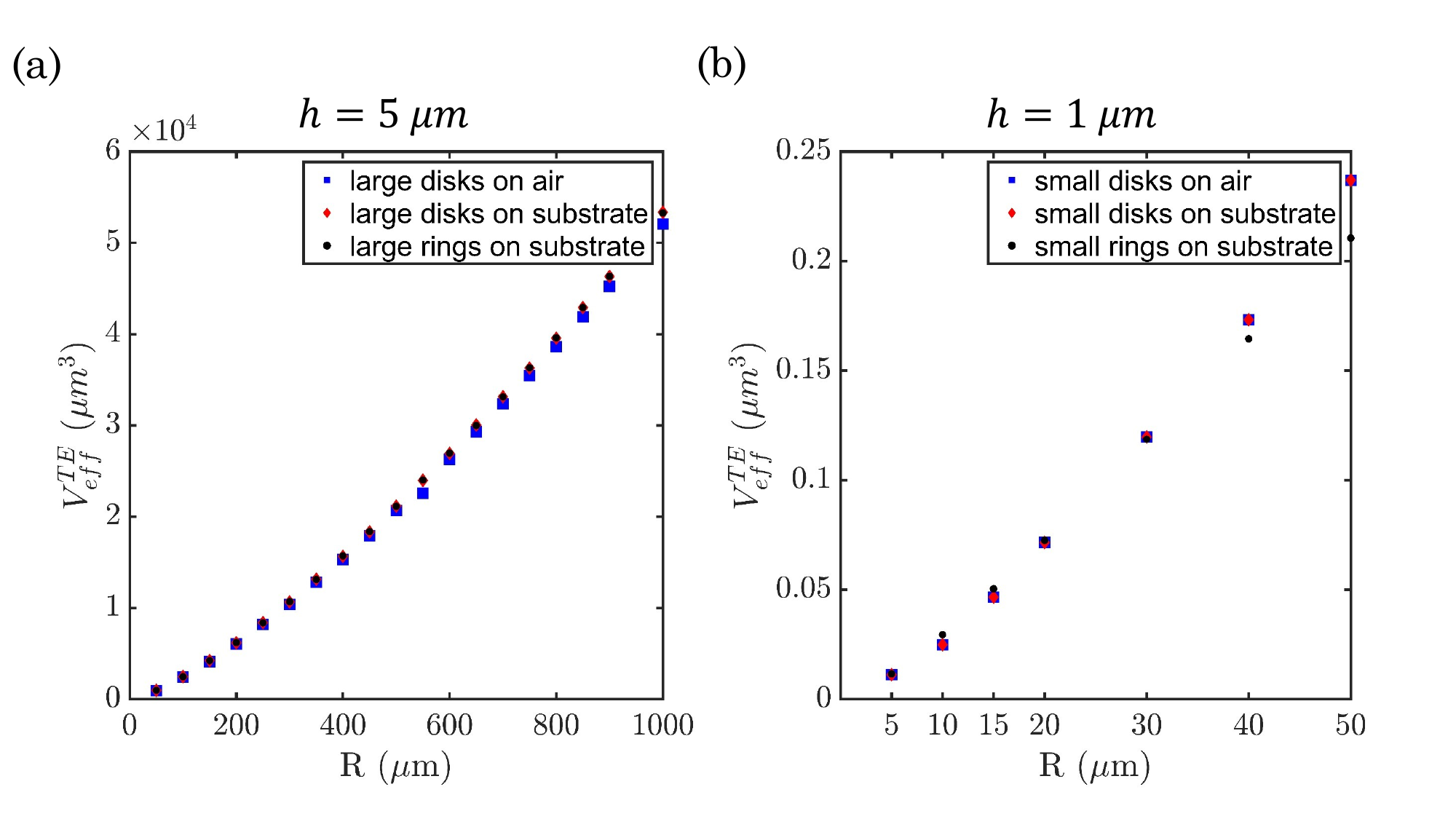} 
\caption{The WGM TE mode volume as a function of the resonator radius for large (a) and small (b) disks/rings.}
\label{fig2: mode volume}
\end{figure*}

\bibliography{optomagnon}

\end{document}